\documentclass{birkmult}
\usepackage[utf8]{inputenc}
\usepackage[T1]{fontenc}
\usepackage{amsmath,amsfonts,amssymb,amsthm}

\newtheorem*{definition}{Definition}

\numberwithin{equation}{section}

\newcommand{\R}{\mathbb{R}}
\newcommand{\C}{\mathbb{C}}

\newcommand{\hyp}[3]{\left(\begin{matrix} #1 \\ #2 \end{matrix};#3 \right)}

\newcommand{\cn}{\chi_{_N}}

\newcommand{\carg}[1]{\chi_{_{#1}}}

\begin{document}
%
%
%
%
%
%
%
%
%

\title{Signal processing, orthogonal polynomials, and Heun equations}

\author{Geoffroy Bergeron}
\address{%
Centre de recherches math\'ematiques, Universit\'e de Montr\'eal,\\
P.O. Box 6128, Centre-ville Station,\\
Montr\'eal, Canada H3C 3J7
}
\email{bergerog@crm.umontreal.ca}

\author{Luc Vinet}
\address{%
Centre de recherches math\'ematiques, Universit\'e de Montr\'eal,\\
P.O. Box 6128, Centre-ville Station,\\
Montr\'eal, Canada H3C 3J7
}
\email{vinet@crm.umontreal.ca}

\author{Alexei Zhedanov}
\address{%
Department of Mathematics, School of Information,\\
Renmin University of China,\\
Beijing 100872, China.
}
\email{zhedanov@yahoo.com}

\date{\today}

\begin{abstract}
A survey of recents advances in the theory of Heun operators is offered. Some of the topics covered include: quadratic algebras and orthogonal polynomials, differential and difference Heun operators associated to Jacobi and Hahn polynomials, connections with time and band limiting problems in signal processing.
\end{abstract} 

\maketitle

\section{Introduction}
This lecture aims to present an introduction to the algebraic approach to Heun equation. To offer some motivation, we shall start with an overview of a central problem in signal treatment, namely that of time and band limiting. Our stepping stone will be the fact that Heun type operators play a central role in this analysis thanks to the work of Landau, Pollack and Slepian \cite{slepian}, see also the nice overview in \cite{timeband}. After reminding ourselves of the standard Heun equation, we shall launch into our forays. We shall recall that all polynomials of the Askey scheme are solutions to bispectral problems and we shall indicate that all their properties can be encoded into quadratic algebras that bear the name of these families. We shall use the Jacobi polynomials as example. We shall then discuss the tridiagonalization procedure designed to move from lower to higher families of polynomials in the Askey hierarchy. This will be illustrated by obtaining the Wilson/Racah polynomials from the Jacobi ones or equivalently by embedding the Racah algebra in the Jacobi algebra. We shall then show that the standard Heun operator can be obtained from the most general tridiagonalization of the hypergeometric (the Jacobi) operator. This will lead us to recognize that an algebraic Heun operator can be associated to each of entries of the Askey tableau. We shall then proceed to identify the Heun operator associated to the Hahn polynomials. It will be seen to provide a difference version of the standard Heun operator. We shall have a look at the algebra this operator forms with the Hahn operator and of its relation to the Racah algebra. We shall then loop the loop by discussing the finite version of the time and band limiting problem and by indicating how the Heun-Hahn operator naturally provides a tridiagonal operator commuting with the non-local limiting operators. We shall conclude with a summary of the lessons we will have learned.

\section{Motivation and background}
\subsection{Time and band limiting}
A central problem in signal processing is that of the optimal reconstruction of a signal from limited observational data. Several physical constraints arise when sampling a signal. We will here focus on those corresponding to a limited time window and to to a cap on the detection of frequencies. Consider a signal represented as a function of time by
$$ f :\R \longrightarrow \R,$$
and suppose $f$ can only be observed for a finite time interval
$$ W = [-T,T]\subset \R.$$
This time limiting can be expressed as multiplication by a step function $W$ defined by
\begin{align*}
\chi_{_W}(t) =
\begin{cases}
1, & \text{if }-T \leq t \leq T,\\
0, & \text{otherwise.}
\end{cases}
\end{align*}
Now, suppose the measurements are limited in their bandwidth. This corresponds to an upper bound on accessible frequencies. Let us express this band limiting as multiplication by a step function  $\chi_{{}_N}$ of the Fourier transform of the signal $f$, where
\begin{align*}
\chi_{{}_N}(n)=
\begin{cases}
1, & \text{if }0 \leq n \leq N,\\
0, & \text{otherwise.}
\end{cases}
\end{align*}
This defines the time limiting operator $\chi_{{}_W}$
\begin{align*}
\chi_{{}_W}: \mathcal{C}(\R) \longrightarrow \mathcal{C}(\R),
\end{align*}
acting by multiplication on functions of time and the band limiting operator $\chi_{{}_N}$
\begin{align*}
\chi_{{}_N} : \mathcal{C}(\R) \longrightarrow \mathcal{C}(\R),
\end{align*}
acting by multiplication on functions of frequencies. Thus, the available data on $f$ is limited to $\chi_{{}_N}\, F\, \chi_{{}_W}\, f$, where $F$ denotes the Fourier transform. The time and band limiting problem consists in the optimal reconstruction of $f$ from the limited available data $\chi_{{}_N}\, F\, \chi_{{}_W}\, f$.

In this context, the best approximation of $f$ requires finding the singular vectors of the operator
\begin{align*}
E = \chi_{_N}\, F \, \chi_{_W} ,
\end{align*}
which amounts to the eigenvalue problems for the following operators
\begin{align*}
E^* E = \chi_{_W} F^{-1} \chi_{_N} F \chi_{_W}, \quad \text{ and } \quad E E^* = \chi_{_N} F \chi_{_W} F^{-1} \chi_{_N}.
\end{align*}
For $F$ the standard Fourier transform, one has
\begin{align}
 \left[ E E^*\, \tilde{f} \right] (l) &= \chi_{_N}\int\limits_{-T}^{T} e^{ilt}\left(\int\limits_{0}^{N} \tilde{f}(k) e^{-ikt} dk\right)dt\nonumber\\
 &= \cn \int\limits_{0}^{N} \tilde{f}(k)\left(\int\limits_{-T}^{T} e^{i(l-k)t} dt \right) dk,\nonumber\\
 &= \int K_T(l,k)\tilde{f}(k)dk,\label{chopper}
 \end{align}
where
\begin{align*}
 K_T(l,k) = \int\limits_{-T}^{T} e^{i(l-k)T}dt = \frac{\sin (l-k)T}{(l-k)},
\end{align*}
which is the integral operator with the well-known \emph{sinc kernel}. It is known, that non local operators such as $E^* E$ have spectra that are not well-suited to numerical analysis. This makes difficult obtaining solutions to the time and band limiting problem. However, a remarkable observation of Landau, Pollak and Slepian \cite{lsp1,lsp2,lsp3,lsp4,lsp5} is that there is a differential operator $D$ with a well-behaved spectrum that commutes with the integral operator $E^* E$. This reduces the time and band limiting problem to the numerically tractable eigenvalue problem of $D$. In the above example, this operator $D$ is a special case of the Heun operator. The algebraic approach presented here will give indications (in the discrete-discrete case in particular) as to why this "miracle" happens.

\subsection{The Heun operator}
Let us first remind ourselves of basic facts regarding the usual Heun operator \cite{heun}. The Heun equation is the Fuchsian differential equation with four regular singularities. The standard form is obtained through homographic transformations by placing the singularities at $x=0,1,d$ and $\infty$ and is given by
\begin{align*}
\frac{d^2}{dx^2} \psi(x) + \left(\frac{\gamma}{x}+\frac{\delta}{x-1}+\frac{\epsilon}{x-d}\right)\frac{d}{dx}\psi(x) + \frac{\alpha\beta x-q}{x(x-1)(x-d)} \psi(x) = 0,
\end{align*}
where
$$\alpha+\beta-\gamma-\delta+1=0,$$
to ensure regularity of the singular point at $x=\infty$. This Heun equation can be written in the form
\begin{align*}
M \psi(x) = \lambda \psi(x)
\end{align*}
with $M$ the Heun operator given by
\begin{align}\label{diffheun}
M = x(x-1)(x-d)\frac{d^2}{dx^2}+(\rho_2 x^2 + \rho_1 x + \rho_0) \frac{d}{dx}+ r_1 x + r_0,
\end{align}
with
\begin{align*}
\quad \rho_2 &= -(\gamma + \delta +\epsilon), & \rho_1 &= (\gamma+\delta)d + \gamma + \epsilon,\\
 \rho_0 &= -\gamma d, &&\\
 r_1 &= -\alpha \beta, & r_0 &= q + \lambda.
\end{align*}
One can observe that $M$ sends any polynomials of degree $n$ to a polynomial of degree $n+1$. Indeed, the Heun operator can be defined as the most general second order differential operator that sends any polynomials of degree $n$ to polynomials of degree $n+1$.

\section{The Askey scheme and bispectral problems}
A pair of linear operators $X$ and $Y$ is said to be bispectral if there is a two-parameter family of common eigenvectors $\psi(x,n)$ such that one has
\begin{align*}
 X \psi(x,n) &= \omega(x) \psi(x,n)\\
 Y \psi(x,n) &= \lambda(n) \psi(x,n),
\end{align*}
with $X$ acting on the variable $n$ and $Y$, on the variable $x$. For the band-time limiting problem associated to sinc kernel, one has the two-parameter family of eigenfunctions given by $ \psi(t,n) = e^{itn}$ with the bispectral pair identified as
\begin{align*}
X &= -\frac{d^2}{dn^2}, & \omega(t) &= t^2,\\
Y &=-\frac{d^2}{dt^2}, & \lambda(n) &= n^2.
\end{align*}
In this case, both operators are differential operators. However, bispectral pairs are realized in terms of various combinations of continuous and discrete operators.
These bispectral problems admit two representations, corresponding to the two spectral parameters $x$ and $n$. As $X$ and $Y$ do not commute, products of these operators must be taken within the same representation for all terms in the product.

A key observation is that each family of hypergeometric polynomials of the Askey scheme defines a bispectral problem. Indeed, these polynomials are the solution to both a recurrence relation and a differential or difference equation. By associating $X$ with the recurrence relation and $Y$ with the differential or difference equation, one forms a bispectral problem as follows. In the $x$-representation, $X$ acts a multiplication by the variable and $Y$ as the differential or difference operator while in the $n$-representation, $X$ acts as a three-term difference operator over $n$ and $Y$ as multiplication by the eigenvalue. The family of common eigenvectors are the orthogonal polynomials.

As a relevant example, consider the (monic) Jacobi polynomials $\hat{P}_n^{(\alpha,\beta)}(x)$ defined as follows \cite{koekoek}
\begin{align*}
\hat{P}_n^{(\alpha,\beta)}(x) = \frac{(-1)^n (\alpha+1)_n}{(\alpha+\beta+n+1)_n}\, {}_2F_1\hyp{-n,n+\alpha+\beta+1}{\alpha+1}{x}.
\end{align*}
These polynomials are the eigenvectors of the hypergeometric operator $D_{x}$ given by
\begin{align}\label{hypergeoop}
D_{x} &\equiv x(x-1)\frac{d^2}{dx^2} + (\alpha+1-(\alpha+\beta+2)x)\frac{d}{dx},
\end{align}
such that
$$D_{x}\, \hat{P}_n^{(\alpha,\beta)}(x) = \lambda_n\, \hat{P}_n^{(\alpha,\beta)}(x),$$
with eigenvalues given by $ \lambda_n = -n(n+\alpha+\beta+1)$. They form an orthogonal set:
\begin{align}\label{jacobiortho}
\int\limits_0^1 \hat{P}_n^{(\alpha,\beta)}(x) \hat{P}_m^{(\alpha,\beta)}(x) x^{\alpha}(1-x)\beta dx = h_n \delta_{n,m},
\end{align}
where
$$ h_n = \frac{\Gamma(\alpha+1)\Gamma(\beta+1)}{\Gamma(\alpha+\beta_2)}u_1 u_2 \cdots u_n. $$
The Jacobi polynomials also satisfy the three-term recurrence relation given by
\begin{align}\label{jacobirecu}
x \hat{P}_n^{(\alpha,\beta)}(x) = \hat{P}_{n+1}^{(\alpha,\beta)}(x) + b_n \hat{P}_n^{(\alpha,\beta)}(x) + u_n \hat{P}_{n-1}^{(\alpha,\beta)}(x),
\end{align}
where
\begin{align*}
  u_n &= \frac{n(n+\alpha)(n+\beta)(n+\alpha+\beta)}{(2n+\alpha+\beta-1)(2n+\alpha+\beta)^2(2n+\alpha+\beta+1)},\\
  b_n &= \frac{1}{2}+ \frac{\alpha^2-\beta^2}{4}\left(\frac{1}{2n+\alpha+\beta} - \frac{1}{2n+\alpha+\beta+2}\right).
\end{align*}
Taking 
\begin{align*}
 X = x, \qquad Y = D_x,
\end{align*}
for the $x$-representation and
\begin{align*}
 X &= T_n^+ + b_n\cdot 1 + u_n T_n^{-1}, \qquad Y = \lambda_n, \qquad \text{where} \qquad T_n^{\pm} f_n = f_{n\pm1},
\end{align*}
for the $n$-representation, the Jacobi polynomials provide a two-parameter set of common eigenvectors of $X$ and $Y$ and hence of the bispectral problem they define. This construction arises similarly for all the orthogonal polynomials in the Askey scheme.

\subsection{An algebraic description}

The properties of the orthogonal polynomials of the Askey scheme can be encoded in an algebra as follows. For any such polynomials, take the $X$ operator to be the multiplication by the variable and the $Y$ operator as the differential or difference equation they satisfy. Consider then the associative algebra generated by $K_1$, $K_2$ and $K_3$ where
\begin{align}\label{polyalgebra}
 K_1 \equiv X, \qquad K_2 \equiv Y, \qquad K_3 \equiv [K_1, K_2].
\end{align}
Upon using these definitions for the generators, one can derive explicitly the commutation relations to obtain that $[K_2,K_3]$ and $[K_3,K_1]$ are quadratic expressions in $K_1$ and $K_2$. Once these relations have been identified, the algebra can be posited abstractly and the properties of the corresponding polynomials follow from representation theory.

Sitting at the top of the Askey scheme, the Wilson and Racah polynomials \cite{koekoek} are the most general ones and the algebra encoding their properties encompasses the other. As the algebraic description is insensitive to truncation, both the Wilson and Racah polynomials are associated to the same algebra. This algebra is known as the Racah-Wilson or \emph{Racah algebra} and is defined \cite{Genest2014} as the associative algebra over $\C$ generated by $\{ K_1, K_2, K_3 \}$ with relations
\begin{align}\label{racahalgebra}
 [K_1,K_2] &= K_3\\
 [K_2, K_3] &= a_1\{K_1,K_2\} + a_2 K_2^2 + b K_2 + c_1 K_1 +d_1 I\\
 [K_3,K_1] &= a_1 K_1^2 + a_2 \{K_1, K_2\} + bK_1 + c_2K_2 + d_2 I,
\end{align}
where $a_1,a_2,b,c_1,c_2,d_1$ and $d_2$ are structure parameters and where $\{A,B\}=AB+BA$ denotes the anti-commutator. One can show that the Jacobi identity is satisfied. The Racah algebra naturally arises in the study of classical orthogonal polynomials but has proved useful in the construction of integrable models and in representation theory \cite{Genest2013,Genest2014}.

Other polynomials of the Askey scheme can be obtained from the Racah or Wilson polynomials by limits and specializations. The associated algebras can be obtained from the Racah algebra in the same way. In particular, the \emph{Jacobi algebra} \cite{1} constitutes one such specialization where $a_1,c_1,d_1,d_2\rightarrow 0$. Indeed, taking
\begin{align}
A_1 &= Y = D_x \equiv x(x-1)\frac{d^2}{dx^2} + (\alpha+1-(\alpha+\beta+2)x)\frac{d}{dx},\nonumber\\
 A_2 &= X =x,\label{jacobialgebradiff}
\end{align}
one finds the following relations for the Jacobi algebra
\begin{align}\label{jacobialgebra}
 [A_1,A_2] &= A_3\\
 [A_2,A_3] &= a_2 A_2^2 + dA_2\\
 [A_3,A_1] &= a_2 \{ A_1, A_2 \} + d A_1 +c_2A_2 +e_2,
\end{align}
where $a_2=2$, $d=-2$, $c_2=-(\alpha+\beta)(\alpha+\beta+2)$ and $e_2=(\alpha+1)(\alpha+\beta)$.

\subsection{Duality}
The bispectrality of the polynomials in the Askey scheme is related to a notion of duality where the variable and the degree are exchanged. In the algebraic description, this corresponds to exchanging the $X$ and $Y$ operator. Let us make details explicit in the finite-dimensional case where the polynomials satisfy both a second order difference equation and a three-term recurrence relation \cite{3}.

In finite dimension, both the $X$ and $Y$ operator will admit a finite eigenbasis. Let us denote the eigenbasis of $X$ by $\{ e_n \}$ and the one of $Y$ by $\{ d_n \}$ for $n=0,1,2,\dots,N$. One first notices that $Y$ will be tridiagonal in the $X$ eigenbasis and likewise for $X$ in the $Y$ eigenbasis. Explicitly, one has
 \begin{align}\label{xybasis}
  X e_n &= \lambda_n e_n, & Y d_n &= \mu_n d_n,\\
  X d_n &= a_{n+1} d_{n+1} + b_n d_n + a_n d_{n-1}, & Y e_n &= \xi_{n+1}e_{n+1} + \eta_{n} e_n + \xi_{n} e_{n-1},\nonumber\\
&&&  \qquad \qquad \qquad n=0,1,\dots,N\nonumber
 \end{align}
where $\{a_n\}$, $\{b_n\}$, $\{\xi_n\}$ and $\{\eta_n\}$ for $n=0,1,\dots,N$ are scalar coefficients. As both the $X$ and $Y$ eigenbases span the same space, one can expand one basis onto the other as follows
\begin{align}\label{expansion}
 e_s = \sum\limits_{n=0}^{N} \sqrt{w_s} \phi_n(\lambda_s)d_n,
\end{align}
where $\phi_n(x)$ are the polynomials associated to the algebra defined by the following recurrence relation
\begin{align*}
 a_{n+1} \phi_{n+1}(x) + b_n \phi_n(x) + a_n \phi_{n-1}(x) = x \phi_{n}(x), \quad \phi_{-1}=0, \quad \phi_0=1,
\end{align*}
which verify the orthogonality relation
\begin{align*}
 \sum\limits_{s=0}^{N} w_s \phi_n(\lambda_s) \phi_m(\lambda_s) = \delta_{n,m},
\end{align*}
so that the reverse expansion is easily seen to be
\begin{align*}
 d_n = \sum\limits_{s=0}^{N} \sqrt{w_s} \phi_n(\lambda_s) e_s.
\end{align*}

Consider now the \emph{dual} set of polynomials $\carg{n}(x)$ defined by the following recurrence relation
\begin{align*}
 \xi_{n+1} \carg{n+1}(x) + \eta_n \carg{n}(x) + \xi_{n} \carg{n-1}(x) = x \carg{n}(x), \quad \carg{-1}=0,\, \carg{0}=1,
\end{align*}
which are orthogonal with respect to the dual weights $\tilde{w}_s:$
\begin{align}\label{dualortho}
 \sum\limits_{s=0}^{N} \tilde{w}_s \carg{n}(\mu_s) \carg{m}(\mu_s) = \delta_{n,m}. 
\end{align}
These dual polynomials provide an alternative expansion of one basis onto the other. One has
\begin{align}\label{dualexpansion}
 d_s = \sum\limits_{n=0}^{N} \sqrt{\tilde{w}_s} \carg{n}(\mu_s) e_n.
\end{align}
One readily verifies this expansion by applying $Y$ to obtain
\begin{align*}
Y d_s &= \sum\limits_{n=0}^{N} \sqrt{\tilde{w}_s} \carg{n}(\mu_s) Y e_n = \sum\limits_{n=0}^{N} \sqrt{\tilde{w}_s} \carg{n}(\mu_s) [\xi_{n+1}e_{n+1} + \eta_{n} e_n + \xi_{n} e_{n-1}]\\
 &= \sum\limits_{n=0}^{N} \sqrt{\tilde{w}_s} [\xi_{n+1} \carg{n+1}(\mu_s) + \eta_n \carg{n}(\mu_s) + \xi_{n} \carg{n-1}(\mu_s)]e_n = \mu_s d_s.
\end{align*}
Using the orthogonality of the polynomials $\{ \carg{n}(\mu_s) \}$ given by \eqref{dualortho}, the expansion \eqref{dualexpansion} is inverted as
\begin{align*}
 e_n = \sum\limits_{s=0}^{N} \sqrt{\tilde{w}_s} \cn(\mu_s) d_s.
\end{align*}
Comparing the above with the first expansion in \eqref{expansion}, knowing the $\{ d_n \}$ to be orthogonal, one obtains
\begin{align}\label{leonardduality}
 \sqrt{w_s} \phi_n(\lambda_s) = \sqrt{\tilde{w}_n} \carg{s}(\mu_n),
\end{align}
a property known as Leonard duality \cite{leonard}, see also \cite{terwilliger} for an introduction to Leonard pairs.

\section{Tridiagonalization of the hypergeometric operator}
Tridiagonalization enables one to construct orthogonal polynomials with more parameters from simpler ones and thus to build a bottom-up characterization of the families of the Askey scheme from this bootstrapping. In particular, properties of the Wilson and Racah polynomials can be found from the tridiagonalization of the hypergeometric operator \cite{1}. Moreover, by considering the most general tridiagonalization, one recovers the complete Heun opertor \cite{2}.

\subsection{The Wilson and Racah polynomials from the Jacobi polynomials}

In the canonical realization of the Jacobi algebra in terms of differential operators presented in \eqref{jacobialgebradiff}, one of the generator is the hypergeometric operator \eqref{hypergeoop} and the other is the difference operator in the degree corresponding to the recurrence relation \eqref{jacobirecu}. We consider the construction of an operator in the algebra which is tridiagonal in the eigenbases of both operators.

Let $Y=D_{x}$ be the hypergeometric operator and $X=x$ be multiplication by the variable. Define $M$ in the Jacobi algebra as follows
\begin{align}\label{mop}
M = \tau_1 XY + \tau_2 YX + \tau_3 X + \tau_{0},
\end{align}
where $\tau_{i}$, $i=0,1,2,3$ are scalar parameters. Knowing that $X$ leads to the three-term recurrence relation of the Jacobi polynomials $\hat{P}_n^{(\alpha,\beta)}(x)$:
\begin{align*}
X\, \hat{P}_n^{(\alpha,\beta)}(x) = x \hat{P}_n^{(\alpha,\beta)}(x) = \hat{P}_{n+1}^{(\alpha,\beta)}(x) + b_n \hat{P}_n^{(\alpha,\beta)}(x) + u_n \hat{P}_{n-1}^{(\alpha,\beta)}(x),
\end{align*}
and is obviously tridiagonal, it is clear from \eqref{mop} that $M$ will also be tridiagonal in the eigenbasis of Y that the Jacobi polynomials form. One has
\begin{align}\label{mtridiag}
M \hat{P}_n^{(\alpha,\beta)}(x) = \xi_{n+1} \hat{P}_{n+1}^{(\alpha,\beta)}(x) + \eta_n \hat{P}_n^{(\alpha,\beta)}(x) + b_n u_n \hat{P}_{n-1}^{(\alpha,\beta)}(x),
\end{align}
where
\begin{align*}
 \xi_n &= \tau_1 \lambda_{n-1} + \tau_2 \lambda_n + \tau_3,\\
 \eta_n &= (\tau_1+ \tau_2) \lambda_n b_n + \tau_3 b_n,\\
 b_n &= \tau_1 \lambda_n + \tau_2 \lambda_{n-1} + \tau_3.
\end{align*}
If $\tau_{1}+\tau_{2}=0$, then $M$ simplifies to
$
M= \tau_1[X,Y] + \tau_3 X, 
$
which is a first order differential operator. In order for $M$ to remain a second order operator, one demands that $\tau_1+\tau_2\neq 0 $. In this case, normalizing $M$ so that $\tau_1+\tau_2 = 1$, one obtains explicitly
\begin{multline}\label{explicitm}
 M = x^2(x-1) \frac{d^2}{dx^2} + x [\alpha+1-2\tau_2-(\alpha+\beta-2\tau_2)x]\frac{d}{dx}\\
 - [\tau_2(\alpha+\beta+2)-\tau_3]x + (\alpha+1)\tau_2+\tau_{0},
\end{multline}
We now construct a basis in which $M$ is diagonal. In the realization \eqref{jacobialgebradiff}, where the algebra acts on functions of $x$, $X$ is multiplication by x and its inverse is defined by
\begin{align*}
X^{-1} : f(x) \longmapsto \frac{1}{x} f(x).
\end{align*}
With this definition, one can invert the expression for $M$ given by \eqref{mop} to obtain
\begin{align}\label{yfromm}
Y = \tau_{1} X^{-1} M + \tau_{2} M X^{-1} + (2 \tau_{1} \tau_{2} - \tau_{0}) X^{-1} - (2 \tau_{1} \tau_{2} + \tau_{3}).
\end{align}
Observing that \eqref{yfromm} has the same structure as \eqref{mop} under the transformation $X \mapsto X^{-1}$, the eigenfunctions of $M$ can be constructed as follows. Introduce the variable $y=1/x$ and conjugate $M$ and $Y$ by a monomial in $y$ to obtain
\begin{align*}
 \tilde{Y} &= y^{\nu-1} Y y^{1-\nu}, & \tilde{M} &= y^{\nu-1} M y^{1-\nu}.
 \end{align*}
Then, by demanding that
\begin{align*}
\tau_{3}=(4+\alpha+\beta-\nu)(\tau_{2}+\nu-1)-\nu\tau_{2},
\end{align*}
the conjugated operators take the following form,
 \begin{align*}
  -\tilde{Y} &= y^2(y-1)\frac{d^2}{dy^2} + y(a_1 y + b_1)\frac{d}{dy} + c_1 y + d_1,\\
  -\tilde{M} &= y(y-1) \frac{d^2}{dy^2} + (a_2 y + b_2) \frac{d}{dy} + d_2,
 \end{align*}
with all the new parameters being simple expressions in terms of $\alpha$, $\beta$, $\tau_{0}$, $\tau_{2}$ and $\nu$. Up to a global sign, one recognizes $\tilde{M}$ as the hypergeometric operator in terms of the variable $y$, while $\tilde{Y}$ is similar to $M$. As the Jacobi polynomials diagonalizes the hypergeometric operator, the eigenvectors satisfying
\begin{align}\label{meigen}
M \psi_{n}(x) = \tilde{\lambda_n} \psi_n(x)
\end{align}
are easily found to be
\begin{align*}
\psi_n(x) &= x^{\nu-1} \hat{P}_n^{(\tilde{\alpha},\tilde{\beta})}\left(1/x \right), & \tilde{\lambda_n} &= n(n+\tilde{\alpha}+\tilde{\beta}+1),\\
\tilde{\beta} &=\beta, & \tilde{\alpha} &= 2(\tau_{2}+\nu)-\alpha-\beta-7.
\end{align*}
It follows from the recurrence relation of the Jacobi polynomials \eqref{jacobirecu} that $X^{-1}$ is tridiagonal in the basis $\psi_{n}(x)$ as it corresponds to multiplication by the variable. Thus, a glance at \eqref{yfromm} confirms that $Y$ is tridiagonal in the $\psi_{n}(x)$ basis.

In order to relate this result with the Wilson and Racah orthogonal polynomials, consider the  expansion of $\psi_{n}(x)$ in terms of $\hat{P}_k^{(\alpha,\beta)}(x)$. One has
\begin{align}\label{ytombasisexp}
\psi_n(x) = \sum\limits_{k=0}^{\infty} G_k(n) \hat{P}_k^{(\alpha,\beta)}(x).
\end{align}
By factoring the expansion coefficients as $G_k(n) = G_0(n) \Xi_k Q_k(n)$, one finds using \eqref{mtridiag} and \eqref{meigen} that, for a unique choice of $\Xi_k$, $Q_k$ satisfies the following three-term recurrence relation
\begin{align*}
\tilde{\lambda_n} Q_k(n) = B_{k} Q_{k+1}(n) + U_k Q_k(n) + F_k Q_{k-1}(n),
\end{align*}
where
\begin{align}\label{wilsonreccoeff}
B_{k} &= u_{k+1} (\tau_1\lambda_{k+1} + \tau_2 \lambda_{k} + \tau_3),\nonumber\\
U_{k} &= \lambda_k b_k + \tau_3 b_k,\\
F_{k} &= \tau_{1} \lambda_{k-1} + \tau_{2} \lambda_{k} + \tau_{3}.\nonumber
\end{align}
The recurrence relation allows to identify the factor $Q_k(n)$ of the expansion coefficient in \eqref{ytombasisexp} as four parameters Wilson polynomials $W_n(x;k_1,k_2,k_3,k_4)$. In this construction, two of these parameters are inherited from the Jacobi polynomials while, after scaling, the tridiagonalisation introduced two free parameters.

The Racah polynomials occur in this setting when a supplementary restriction is introduced. Indeed, a glance at \eqref{explicitm} shows that the generic $M$ operator maps polynomials of degree $n$ into polynomials of degree $n+1$. However, one can see from \eqref{mtridiag} that if
\begin{align*}
\xi_{N+1} = \tau_1\lambda_{N} + \tau_2 \lambda_{N+1} + \tau_3 = 0,
\end{align*}
both $Y$ and $M$ preserve the space of polynomials of degree less or equal to $N$. This truncation condition is satisfied when $\nu=N+1=2-2\tau_{2}$. In this case, the eigenvectors of $M$ are
\begin{align*}
 \psi_n(x) &= x^N \hat{P}_n^{(N-\alpha-\beta-4,\beta)}(1/x),
 \end{align*}
 which are manifestly polynomials of degree $N-n$. One then considers again the expansion of the basis element $\psi_{n}(x)$ into $\hat{P}_k^{(\alpha,\beta)}(x)$ to obtain
 \begin{align*}
 \psi_n(x) &= \sum\limits_{k=0}^{N}R_{n,k} \hat{P}_k^{(\alpha,\beta)}(x),
\end{align*}
where the expansion coefficients $R_{n,k}$ can be shown to be given in terms of the Racah polynomials. Using the orthogonality of the Jacobi polynomials given in \eqref{jacobiortho}, one obtains
$$ R_{n,k} h_k = \int\limits_{0}^{1} \psi_n(x) \hat{P}_k^{(\alpha,\beta)}(x)\, x^{\alpha}(1-x)^{\beta} dx, $$
an analog of the Jacobi-Fourier transform of Koornwinder \cite{koornwinder}, giving an integral representation of the Racah polynomials.

It was stated earlier that the properties of the orthogonal polynomials in the Askey scheme are encoded in their associated algebras. This can be seen from the construction of the Wilson and Racah polynomials from the Jacobi polynomials by the tridiagonalization procedure which corresponds algebraically to an embedding of the Racah algebra in the Jacobi algebra. This is explicitly given by
\begin{align}\label{racahinjacobi}
 K_1 &= A_1, & K_2 &= \tau_1 A_2 A_1 + \tau_2 A_1 A_2 + \tau_3 A_2,
\end{align}
where $A_{1}$, $A_{2}$ are the Jacobi algebra generators as in \eqref{jacobialgebradiff}. One shows that $K_{1}$ and $K_{2}$ as defined in \eqref{racahinjacobi} verify the relations \eqref{racahalgebra} of the Racah algebra assuming that $A_{1}$ and $A_{2}$ verify those of the Jacobi relations as given in \eqref{jacobialgebra}. Thus, the embedding \eqref{racahinjacobi} encodes the tridiagonalization result abstractly.

The tridiagonalisation \eqref{mop} used to derive higher polynomials from the Jacobi polynomials is not the most general tridiagonal operator that can be constructed from the Jacobi algebra generators. Indeed, consider the addition in \eqref{mop} of a linear term in $Y$, given by \eqref{hypergeoop}:
\begin{align}\label{mop2}
M = \tau_1 XY +\tau_2 YX +\tau_3 X + \tau_4 Y + \tau_{0}.
\end{align}
It is straightforward to see that $M$ as given by \eqref{mop2} is equal to the Heun operator \eqref{diffheun}. Expressed as in \eqref{mop2}, the Heun operator is manifestly tridiagonal on the Jacobi polynomials, which offers a simple derivation of a classical result. For the finite dimensional situation see \cite{nomura}.

\section{The Algebraic Heun operator}

The emergence of the standard Heun operator from the tridiagonalization of the hypergeometric operator suggests that Heun-type operators can be associated to bispectral problems. In particular, knowing all polynomials in the Askey scheme to define bispectral problems, there should be Heun-like operators associated to each of these families of polynomials. Guided by this observation, consider a set of polynomials in the Askey scheme and let $X$ and $Y$ be the generators of the associated algebra as in \eqref{polyalgebra}. As before, $X$ is the recurrence operator and $Y$, the difference or differential operator. The corresponding Heun-type operator $W$ is defined as
\begin{align}\label{wop}
W = \tau_1 XY + \tau_2 YX + \tau_3 X + \tau_4 Y + \tau_0,
\end{align}
and will be referred to as an \emph{algebraic Heun operator} \cite{3}. The operator $W$ associated to a polynomial family will have features similar to those of the standard Heun operator which arises in the context of the Jacobi polynomials. To illustrate this, a construction that parallels the one made for the Jacobi polynomials is presented

\subsection{A discrete analog of the Heun operator}

The standard Heun operator can be defined as the most general degree increasing second order differential operator. In analogy with this, one defines the \emph{difference Heun operator} as:
\begin{definition}[Difference Heun operator]
The difference Heun operator is the most general second order difference operator on a uniform grid which sends polynomials of degree $n$ to polynomials of degree $n+1$.
\end{definition}
We now obtain an explicit expression for the difference Heun operator on the finite grid $G=\{0,1,\dots,N\}$. Let $T^{\pm}$ be shift operators defined by
\begin{align}\label{shiftop}
T^{\pm} f(x) &= f(x\pm1),
\end{align}
and take $W$ to be a generic second order difference operator with
\begin{align}\label{wdiff}
W &= A_1(x) T^{+} + A_2(x) T^{-} + A_0(x) I.
\end{align}
By demanding that $W$ acting on $1$, $x$ and $x^{2}$ yields polynomials of one degree higher, one obtains that
\begin{align}\label{wdiff2}
A_0(x) = \tilde{\pi}_1(x)-\tilde{\pi}_3(x), \qquad A_1(x) = \frac{\tilde{\pi}_3(x)-\tilde{\pi}_2(x)}{2}, \qquad A_2(x) = \frac{\tilde{\pi}_3(x)+\tilde{\pi}_2(x)}{2},
\end{align}
where the $\tilde{\pi}_{i}(x)$ are arbitrary polynomials of degree $i$ for $i=1,2,3$. Thus, in general, $A_{i}(x)$ for $i=0,1,2$ are third degree polynomials with $A_{1}(x)$ and $A_{2}(x)$ having the same leading coefficient. Moreover, the restriction of the action of $W$ to the finite grid $G$ implies that $A_{1}$ has $(x-N)$ as a factor and $A_{2}$ has $x$ as a factor. Hence, one has
\begin{align*}
A_1(x) &= (x-N)(\kappa x^2 + \mu_1 x +\mu_0), \\
A_2(x) &= x(\kappa x^2 + \nu_1 x + \nu_0), \\
A_0 (x) &= -A_1(x) - A_2(x) + r_1 x + r_0,
\end{align*}
for $\mu_{0}, \mu_{1},\nu_{0},\nu_{1},r_{0},r_{1}$ and $\kappa$ arbitrary parameters. Then, it is easy to see that
\begin{align*}
W[x^{n}] = \sigma_n\, x^{n+1}+O(x^{n}),
\end{align*}
for a certain $\sigma_{n}$ depending on the parameters. We shall see next that this difference Heun operator coincides with the algebraic Heun operator associated to the Hahn algebra.

\subsection{The algebraic Heun operator of the Hahn type}
The Hahn polynomials $P_n$ are orthogonal polynomials belonging to the Askey scheme. As such, an algebra encoding their properties is obtained as a specialization of the Racah algebra \eqref{racahalgebra} by taking $a_{2}\rightarrow 0$. One obtains the Hahn algebra, generated by $\{ K_{1}, K_{2}, K_{3} \}$ with the following relations
\begin{align}\label{hahnalgebra}
 [K_1,K_2] &= K_3,\nonumber\\
 [K_2,K_3] &= a\{ K_1,k_2 \} + b K_2 + c_1 K_1 + d_1 I,\nonumber\\
 [K_3,K_1] &= a K_1^2 + b K_1 + c_2 K_2 + d_2 I.
\end{align}
A natural realization of the Hahn algebra is given in terms of the bispectral operators associated to the Hahn polynomials $P_n$, namely,
\begin{align}\label{hahnbisreal}
 X &= K_1 = x,\\
 Y &= K_2 = B(x) T^{+} + D(x) T^{-} - (B(x)- D(x)) I,\nonumber
\end{align}
with
\begin{align*}
 B(x) = (x-N)(x+ \alpha +1), \qquad\qquad D(x) = x(x-\beta-N-1),
\end{align*}
and where $T^{\pm}$ is as in \eqref{shiftop}. The action of $Y$ is diagonal in the basis given by the Hahn polynomial $P_{n}$ and is
\begin{align*}
Y P_n(x) = \lambda P_n(x), \qquad \lambda_n = n(n+\alpha+\beta +1).
\end{align*}
One checks that $X$ and $Y$ satisfy the Hahn algebra relations \eqref{hahnalgebra} with the structure constants expressed in terms of $\alpha$, $\beta$ and $N$.

Upon identifying the algebra associated to the Hahn polynomials, one can introduce the algebraic Heun operator $W$ of the Hahn type \cite{4} using the generic definition \eqref{wop}. In this realization, one finds that $W$ can be written as
\begin{align*}
W = A_1(x) T^{+} + A_2(x) T^{-} + A_0(x) I,
\end{align*}
where
\begin{align*}
 A_1(x) &= (x-N)(x+\alpha+1)((\tau_1+\tau_2)x+\tau_2+\tau_4),\\
 A_2(x) &= x(x-\beta-N-1)((\tau_1+\tau_2)x + \tau_4 - \tau_2),\\
 A_0(x) &= -A_1(x) -A_2(x) + ((\alpha+\beta+2)\tau_2 + \tau_3)x + \tau_0 - N(\alpha+1)\tau_2.
\end{align*}
As announced, the operator defined above coincides, upon identification of parameters, with the difference Heun operator $W$ given in \eqref{wdiff} and \eqref{wdiff2} and defined through its degree raising action on polynomials. That the difference Heun operator is tridiagonal on the Hahn polynomials then follows as a direct result. This parallels the construction in the Jacobi algebra that led to a simple proof of the standard Heun operator being tridiagonal on the Jacobi polynomials. Moreover, in the limit $N \rightarrow\infty$, the difference Heun operator $W$ goes to the standard Heun operator, which further supports the appropriateness of the abstract definition \eqref{wop} for the algebraic Heun operator.

To conclude this algebraic analysis, let us consider the algebra generated by $Y$ and $W$ in the context of the Hahn algebra. By introducing a third generator given by $[W,Y]$ and using the relation of the Hahn algebra in \eqref{hahnalgebra}, one finds that the algebra thus generated closes as a cubic algebra with relations given by
\begin{align*}
 [Y,[W,Y]] &= g_1 Y^2 + g_2 \{ Y,W \} + g_3 Y + g_4 W + g_5 I,\\
 [[W,Y],W] &= e_1 Y^2 + e_2 Y^3 + g_2 W^2 + g_1 \{ Y,W \}  + g_3 W + g_6 Y + g_7 I,
\end{align*}
where the structure constants depend on the parameters of the Hahn polynomials and the parameters of the tridiagonalization \eqref{wop}. One can recognize the above as a generalization of the Racah algebra \eqref{racahalgebra} with the following two additional terms:
\begin{align*}
e_1 Y^2 + e_2 Y^3.
\end{align*}
The conditions for these terms to vanish are given by
\begin{align*}
\tau_1+\tau_2 =0, \qquad \tau_2 \pm \tau_4 = 0.
\end{align*}
When these equalities are satisfied, the operator $W$ simplifies to $W_{+}$ or $W_{-}$ with
\begin{align*}
W_{\pm} = \pm \frac{1}{2} [X,Y] \pm \gamma X - \frac{Y}{2} \pm \epsilon I.
\end{align*}
Moreover, any pair from the set $\{ Y, W_{+}, W_{-} \}$ satisfies the Racah algebra relations given by \eqref{racahalgebra}. Thus, the choice of a pair of operators specifies an embedding of the Racah algebra in the Hahn algebra, which is analogous to the embedding given in \eqref{racahinjacobi}. These embeddings encode abstractly the construction of the Racah polynomials starting from the Hahn polynomials and provide another example where higher polynomials are constructed from simpler ones.

\section{Application to time and band limiting}
We now return to the problem of time and band limiting. Consider a finite dimensional bispectral problem as the one associated to the Hahn polynomials. Denote by $\{ e_{n} \}$ and $\{ d_{n} \}$ for $n=1,2,\dots,N$ the two eigenbases of this bispectral problem such that
\begin{align*}
X:\{ e_{n}\} &\rightarrow \{ e_{n}  \}, & X e_n &= \lambda_n e_n,\\
Y:\{ d_{n} \} &\rightarrow \{ d_{n} \}, & Y d_n &= \mu_n d_n.
\end{align*}
In this context, $X$ can be thought of  being associated to discrete time and $Y$ to frequencies. Suppose now that the spectrum of both $X$ and $Y$ are restricted. These restrictions can be modelled as limiting operators in the form of two projections $\pi_{1}$ and $\pi_{2}$ given by
\begin{align}\label{pis}
 \pi_1 e_n &= \begin{cases}
              e_n & \text{if }n\leq J_1,\\
	      0 & \text{if }n> J_1,
             \end{cases}
&
\pi_2 d_n &= \begin{cases}
             d_n & \text{if }n\leq J_2,\\
	     0 & \text{if }n > J_2,
            \end{cases}\\
 \pi_1^2 &= \pi_1, & \pi_2^2 = \pi_2,\nonumber
\end{align}
Simultaneous restrictions on the eigensubspaces of $X$ and $Y$ accessible to sampling lead to the two limiting operators
\begin{align*}
V_1 = \pi_1 \pi_2 \pi_1 = E_1 E_2, \quad V_2 = \pi_2\pi_1\pi_2 = E_2 E_1,
\end{align*}
with
\begin{align*}
 E_1 = \pi_1 \pi_2, \qquad E_2 = \pi_2 \pi_1.
\end{align*}
Here, the limiting operator $V_{1}$ and $V_{2}$ are symmetric and are diagonalizable. A few limit cases are simple. When there are no restriction, $J_{1}=J_{2}=N$, in which case $V_{1}=V_{2}=I$. If the restriction is on only one of the spectra, for instance if $J_{2}=N$, then $V_1=V_2=\pi_1$ having $J_{1}+1$ unit eigenvalues and the other $N-J_{1}$ equal to zero. However, the case where $J_{1}$ and $J_{2}$ are arbitrary is much more complicated.

In the generic case, the eigenbasis expansions \eqref{expansion} and \eqref{dualexpansion} can be used to evaluate the action of $\pi_{2}$ on an eigenvector of $X$. One has,
\begin{align*}
\pi_2 e_n &= \sum\limits_{s=0}^{J_2} \sqrt{w_n} \phi_s(\lambda_n) d_s = \sum\limits_{s=0}^{J_2} \sum\limits_{t=0}^{N} \sqrt{w_n \tilde{w}_s}\phi_s(\lambda_n) \carg{t}(\mu_s) e_t.
\end{align*}
Similarly, one can evaluate the action of $\pi_{1}$ on eigenvectors of $Y$ and obtain
\begin{align}\label{discreteconv}
V_1 e_n &= \pi_1 \pi_2 \pi_1 e_n = \sum\limits_{t=0}^{J_1}\sum\limits_{s=0}^{J_2} \sqrt{w_n \tilde{w}_s} \phi_s(\lambda_n) \carg{t}(\mu_s)e_t = \sum\limits_{t=0}^{J_1} K_{t,n} e_t,
\end{align}
with
\begin{align}\label{discretekernel}
K_{t,n} &= \sum\limits_{s=0}^{J_2} \sqrt{w_n \tilde{w}_s} \phi_s(\lambda_n) \carg{t}(\mu_s)\nonumber\\
&= \sum\limits_{s=0}^{J_2} \sqrt{w_n w_t} \phi_s(\lambda_n) \phi_s(\lambda_t)\\
&= \sum\limits_{s=0}^{J_2} \sqrt{\tilde{w}_s} \carg{n}(\mu_s) \carg{t}(\mu_s),\nonumber
\end{align}
where the Leonard duality relation \eqref{leonardduality} has been used to obtain the last two equalities. The operator $V_{1}$ in \eqref{discreteconv} is the discrete analog of the integral operator \eqref{chopper} that restricts both in time and frequency, with \eqref{discretekernel} being the discrete kernel. As in the initial continuous case, $V_{1}$ and $V_{2}$ are non-local operator and the problem of finding their eigenvectors is numerically difficult. However, if there exists a tridiagonal matrix $M$ that commutes with both $V_{1}$ and $V_{2}$, then $M$ would admit eigenvectors that are shared with $V_{1}$ and $V_{2}$. This renders the discrete time and band limiting problem well controlled. In this context, the tridiagonal matrix $M$ is the discrete analog of a second order differential operator and plays the role of the differential operator found by Landau, Pollak and Slepian for the continuous time and band limiting problem. 

Tridiagonal matrices that commute with the limiting operators $\pi_{1}$ and $\pi_{2}$ in \eqref{pis} will also commute with $V_{1}$ and $V_{2}$. One then wants to find for an $M$ such that
\begin{align}\label{star}
[M,\pi_1] = [M,\pi_2]=0.
\end{align}
Taking $M$ to be an algebraic Heun operator with
\begin{align*}
M = \tau_1 XY + \tau_2 YX + \tau_3 X + \tau_4 Y,
\end{align*}
and using \eqref{star}, one finds the following conditions
\begin{align*}
 \tau_2 = \tau_1, \qquad \tau_1(\lambda_{J_1} + \lambda_{J_1+1}) + \tau_4 &=0, \qquad \tau_1(\mu_{J_2}+\mu_{J_2+1}) + \tau_3 =0.
\end{align*}
Except for the Bannai-Ito spectrum, it is always possible to find $\tau_{3}$ and $\tau_{4}$ satisfying the above \cite{3}, see also \cite{perline}. Hence, the algebraic Heun operator provides the commuting operator that enables efficient solutions to the time and band limiting.

\section*{Conclusion}
This lecture has offered an introduction to the concept of algebraic Heun operators and its applications. This construct stems from the observation that the standard Heun operator can be obtained from the tridiagonalization of the hypergeometric operator.The key idea is to focus on operators that are bilinear in the generators of the quadratic algebras associated to orthogonal polynomials. The Heun type operators obtained in this algebraic fashion, coincide with those arising from the definition that has Heun operators raising by one the degree of arbitrary polynomials. This has been illustrated for the discrete Heun operator in its connection to the Hahn polynomials. This notion of algebraic Heun operator tied to bispectral problems has  moreover been seen to shed light on the occurence of commuting operators in band and time limiting analyses. The exploration of these algebraic Heun operators and the associated algebras has just begun \cite{4,5,6} but the results found so far let us believe that it could lead to significant new advances.

\section*{Acknowledgements}
One of us (L.V.) is very grateful to Mama Foupouagnigni, Wolfram Koepf, AIMS (Cameroun) and the Volkswagen Stiftung for the opportunity to lecture in Douala. G.B. benefitted from a NSERC postgraduate scholarship. The research of L.V. is supported by a NSERC discovery grant and that of A.Z. by the National Science Foundation of China (Grant No. 11711015).

\end{document}